\documentclass{aa}  
\usepackage{graphicx}
\usepackage[varg]{txfonts}
\usepackage[svgnames]{xcolor}
\usepackage[colorlinks=true,linkcolor=blue,urlcolor=blue,citecolor=blue]{hyperref}

\usepackage{natbib}
\bibpunct{(}{)}{;}{a}{}{,}

\defcitealias{Bestenlehner2020}{Bes20}
\defcitealias{Vink2001}{Vin01}

\newcommand{\Bes}{\citetalias{Bestenlehner2020}}
\newcommand{\Vin}{\citetalias{Vink2001}}

\usepackage{multirow}

\begin{document} 
\newcommand{\add}[1]{{#1}}
\newcommand{\addr}[1]{{#1}}

   \title{Impact of main-sequence mass loss on the appearance, structure and evolution of Wolf-Rayet stars}

   \author{J. Josiek\inst{\ref{inst:ari},\ref{inst:obsge}}
          \and
          S. Ekström\inst{\ref{inst:obsge}}
          \and 
          A.A.C. Sander\inst{\ref{inst:ari}}
          }

   \institute{Zentrum f{\"u}r Astronomie der Universit{\"a}t Heidelberg, Astronomisches Rechen-Institut, M{\"o}nchhofstr. 12-14, 69120 Heidelberg, \\Germany\label{inst:ari}\\
             \email{joris.josiek@uni-heidelberg.de}
             \and
   Department of Astronomy, University of Geneva, Chemin Pegasi 51, 1290 Versoix, Switzerland\label{inst:obsge}
             }

   \date{Received 19 January 2024 / Accepted 22 April 2024}

  \abstract{Stellar winds are one of the most important drivers of massive star evolution and a vital source of chemical, mechanical, and radiative feedback on the galactic scale. Despite its significance, mass loss remains a major uncertainty in stellar evolution models. In particular, the interdependencies of different approaches with subsequent evolutionary stages and predicted observable phenomena are far from being systematically understood.}{In this study, we examine the impact of main sequence mass loss on the structure of massive stars throughout their entire evolution. A particular focus is placed on the consequences for entering the Wolf-Rayet (WR) regime and the subsequent evolution.}{Using the Geneva stellar evolution code, we compute grids of single, non-rotating stellar models at solar and \addr{Large Magellanic Cloud} (LMC) metallicity of initial masses between 20 and 120 solar masses, with two representative prescriptions for high and low main sequence mass loss.}{We obtain detailed numerical predictions regarding the structure and evolution of massive stars, and infer the role of main sequence mass loss by comparison of the mass-loss rate prescriptions. We present implications for the overall evolutionary trajectory, including the evolution of WR stars, as well as the effect on stellar yields and stellar populations.}{Mass loss during the main sequence plays an important role due to its ability to affect the sequence and duration of all subsequent phases. We identify several distinct evolutionary paths for massive stars which are signiﬁcantly influenced by the chosen main sequence mass-loss description. \addr{We also discuss the impact of uncertainties other than mass loss on the evolution, in particular those relating to convection.} We further demonstrate that not just the total mass loss, but the specific mass-loss history throughout a star's life is a crucial determinant of many aspects, such as the resulting stellar yields.}

              
   \maketitle
%

\section{Introduction}

Through their powerful winds and high ionizing flux, massive stars play a vital role in shaping the physical and chemical structure of their surroundings \citep{Krause2013}. Unfortunately, massive stars are greatly outnumbered by their low-mass counterparts, making them much more challenging to study. Some phenomena related to massive stars are so rare that we only have a handful of observations of them. For instance, the number of Wolf-Rayet (WR) stars discovered to date is counted in the hundreds \citep{Massey2014, Mauerhan2011, Rosslowe2015, Neugent2019}, an incredibly small number compared to the billions of stars we know in total.

One of the most important ingredients in stellar models is their mass-loss rate. Within the course of their lives, single massive stars can eject more than half of their own initial mass in the form of stellar winds, which not only transfer energy and momentum into the interstellar medium (ISM), but also feed it with new gas that has been chemically enriched with products of nucleosynthesis. Through mass loss, stars also change their own appearance as they progressively remove layers from their surface. When modeling the evolution of stars, this mass loss cannot be predicted within the usual physical framework of the structure models, and thus needs to be taken into account in a prescribed way. Numerous different methods have been proposed to measure, calculate and predict the mass-loss rate of stars, each with their own caveats and uncertainties. For this reason, it is important to consider a wide range of possible mass loss scenarios when attempting to explain stellar evolution as completely as possible.

While on the main sequence (MS), massive stars appear as hot OB-type stars with surface temperatures between 10\,000--50\,000\,K. During this phase, stars lose mass mainly as a result of line-driven winds, i.e.\ through the absorption of photons in specific transitions of atoms in the atmosphere \citep[see, e.g.,][for a comprehensive review]{Vink2022}.  In this work, we consider two recipes which can be used to compute the mass-loss rates in this phase and study their impact on the later phases of evolution.

The MS is \add{arguably} the most \add{inconspicuous} stage of evolution, \add{marked by slow processing of hydrogen in the core, and} \add{comparatively weak} mass loss \add{ from the surface. It} is therefore often pushed into the background in favor of studying phases of stronger and less certain mass-loss rates such as the red supergiant (RSG) phase \citep[e.g.,][]{Massey2023, Beasor2018} or the WR phase \citep[e.g.,][]{Sander2020,Sander2022}. However, there is up to one order of magnitude of discrepancy between different mass-loss descriptions for OB stars \citep[e.g.,][]{Vink2001,Brands2022,Bjoerklund2023}, which is large enough to introduce major uncertainties into their evolution, even without considering the uncertainties of post-MS mass loss. The goal of this paper is to perform a systematic study on the effects of MS mass loss on the evolution of massive stars. For this, we compute grids of evolutionary models using two different prescriptions for the MS mass-loss rate and compare the outcome.

In Sect.\,\ref{sec:mdot-desc}, we review the two \add{MS} mass-loss prescriptions \add{compared} in this study and \add{briefly summarize the mass-loss prescriptions used in the post-MS phases of evolution.} \add{We} then present the parameters of the model grid in Sect.\,\ref{sec:modeling}. In Sect.\,\ref{sec:early-evolution} we present a preliminary characterization of the mass-loss rates, effects on the evolution during the MS, as well as consequences \add{for} stellar yields. Sect.\,\ref{sec:WR-evol} is dedicated to the formation, structure, and evolution of WR stars as a key phase of evolution for many massive stars. \add{In Sect.\,\ref{sec:populations},} we will \add{present} the impact of MS mass loss on the evolutionary endpoint\add{s of the models} as possible SN progenitors \add{as well as populations of massive stars as a whole}. Finally, we will discuss the findings in Sect.\,\ref{sec:discussion} and summarize the results in Sect.\,\ref{sec:conclusion}.


\section{Mass loss descriptions}
\label{sec:mdot-desc}

\subsection{The main sequence}

In this work, we employ two mass-loss prescriptions that will act as prototypes for \add{stronger} and \add{weaker} mass loss on the MS. The de-facto standard MS mass-loss prescription that is used in most of the current grids of evolutionary models and population synthesis codes is the one introduced by \citet[][hereafter referred to as \Vin]{Vink2001}. This recipe is distilled from a set of numerical stellar wind models and provides the mass-loss rate as a function of the global stellar parameters, split over multiple effective temperature domains. When the temperature transitions into a different domain, the wind models predict the recombination or ionization of iron, causing a significant change in the available opacities to drive the wind, which in the models leads to a discontinuity in the trends for the predicted mass-loss rates. Motivated by the bi-stable wind solution found by \citet{Pauldrach1990} for P\,Cygni and the jump in the observed ratio between terminal and escape velocity by \citet{Lamers1995}, these discontinuities are referred to as \textit{bistability jumps}. The existence of the jumps in $\dot{M}$ and their magnitude are an active topic of theoretical and empirical research \citep[e.g.,][]{Petrov2016,Bjoerklund2023,Bernini2023,Krticka2021,Krticka2024}. 

In our models implementing the \Vin{} mass-loss prescription, we account for both the \ion{Fe}{IV/III} as well as the \ion{Fe}{III/II} jumps.
First, the so-called ``characteristic density'' $\langle\rho\rangle$\add{, defined as the density of the wind at 50\% terminal velocity,} is calculated via Eq.\,(23) in \Vin, i.e.,
\begin{equation}
    \log\langle\rho\rangle = -14.94 + 0.85\log (Z/Z_\odot) + 3.2\,\Gamma_\mathrm{e}
\end{equation}
with $\Gamma_\mathrm{e}$ denoting the electron-scattering Eddington parameter. This parameter encodes the ratio between $L$ and $M$ and thus the ratio between the outward force caused by radiation pressure and the inward pull by gravity, which is inherently linked to radiation-driven mass loss. 
From $\langle\rho\rangle$ we then obtain the location of the bistability jumps with
\begin{equation}
    T_\mathrm{eff,jump1}/\mathrm{K} = 61\,200 + 2\,590 \,\log\langle\rho\rangle\text{,}
\end{equation}
and
\begin{equation}
    T_\mathrm{eff,jump2}/\mathrm{K} = 100\,000 + 6\,000 \,\log\langle\rho\rangle\text{.}
\end{equation}
For the hot domain, i.e.\ $T_\mathrm{eff}>T_\mathrm{eff,jump1}$, the mass-loss rate is then determined as
\begin{equation}
\begin{split}
        \log\dot{M} &= -6.697 \\
        &+ 2.194\log\left(\frac{L}{10^5L_\odot}\right) \\
        &- 1.313\log\left(\frac{M}{30\,M_\odot}\right) \\
        &- 1.226\log\left(\frac{v_\infty/v_\mathrm{esc}}{2.0}\right) \\
        &+ 0.933\log\left(\frac{T_\mathrm{eff}}{40\,000\,\mathrm{K}}\right) \\
        &-10.92\left[\log\left(\frac{T_\mathrm{eff}}{40\,000\,\mathrm{K}}\right)\right]^2 \\
        &+ 0.85\log\left(\frac{Z}{Z_\odot}\right)\text{,}
\end{split}
\end{equation}
with $v_\infty/v_\mathrm{esc}=2.6$.
For the intermediate domain, i.e.\ $T_\mathrm{eff,jump1}>T_\mathrm{eff}>T_\mathrm{eff,jump2}$, we use
\begin{equation}
\label{eq:midrange-vinkrecipe}
\begin{split}
        \log\dot{M} &= -6.688 \\
        &+ 2.21\log\left(\frac{L}{10^5L_\odot}\right) \\
        &- 1.339\log\left(\frac{M}{30\,M_\odot}\right) \\
        &- 1.601\log\left(\frac{v_\infty/v_\mathrm{esc}}{2.0}\right) \\
        &+ 1.07\log\left(\frac{T_\mathrm{eff}}{20\,000\,\mathrm{K}}\right) \\
        &+ 0.85\log\left(\frac{Z}{Z_\odot}\right)\text{,}
\end{split}
\end{equation}
with $v_\infty/v_\mathrm{esc}=1.3$.
For the cool temperature domain, i.e.\ $T_\mathrm{eff}<T_\mathrm{eff,jump2}$, we also use Eq.\,\eqref{eq:midrange-vinkrecipe}, but increase the leading constant to $-5.99$ and set $v_\infty/v_\mathrm{esc}=0.7$. The unit of $\dot{M}$ is as usual implied to be $M_\odot\,\text{yr}^{-1}$, which will be valid throughout this paper without explicitly stating so.

Alternatively to directly modeling the winds, there are also empirical and semi-empirical descriptions of mass loss. Based on the wind theory by \citet{Castor1975} combined with the Eddington stellar model, \citet[][hereafter known as \Bes]{Bestenlehner2020} suggested a description of the mass-loss rate depending mainly on $\Gamma_\mathrm{e}$. \Bes{} proposed a mass-loss recipe in the form

\begin{equation}
    \log\dot{M} = \log\dot{M}_0 + \left(\frac{1}{\alpha} + 0.5\right)\log (\Gamma_\text{e}) - \left(\frac{1}{\alpha} +1\right)\log(1-\Gamma_\text{e})\text{,}
\end{equation}
where $\alpha$ is a force-multiplier parameter used to account for the effect of line opacities, and $\dot{M}_0$ is a global scale factor. This recipe intends to connect the regime of stronger, optically thick winds of very massive WNh stars \citep[e.g.,][]{Vink2011,Bestenlehner2020b,Sabhahit2022} with the weaker optically thin winds of OB stars. Specifically, for thin winds (where $\Gamma_\mathrm{e}\ll 1$), the $\log(\Gamma_\mathrm{e})$-term dominates, whereas for thick winds (where $\Gamma_\mathrm{e}\rightarrow 1$) the $\log(1-\Gamma_\mathrm{e})$-term is dominant. At the transition point between the two regimes, where both $\Gamma_\mathrm{e}$-terms cancel each other, the mass-loss rate is $\dot{M}_0$. As a semi-empirical recipe, both of its parameters $\dot{M}_0$ and $\alpha$ must be calibrated on external data, as was done for example by \citet{Bestenlehner2020b} and \citet{Brands2022} on the R136 cluster in the Large Magellanic Cloud (LMC). In this study, we do not aim for a detailed analysis of individual mass-loss recipes, but employ \Bes{} as a prototype of considerably \add{weaker} MS mass loss compared to \Vin{}. Thus, we use the values of $\log\dot{M}_0=-5.19$ and $\alpha=0.456$ from \citet{Brands2022} for all models computed using the \Bes{} recipe, including those at Galactic metallicity, noting that the mass loss in these models will be underestimated when using LMC-calibrated data.

\subsection{Validity domains of the mass-loss prescriptions}

Due to the changing physical conditions in a star's atmosphere, the mechanisms that drive its mass loss vary considerably throughout the course of its evolution. Therefore, the implementation of a single mass loss description must be restricted to the domain where it is applicable based on the underlying physical assumptions, and multiple different mass loss descriptions are used to cover the entire evolution of a star.

For the \Vin{} models, we apply this mass-loss rate in the same domain defined in previous GENEC models \citep{Ekstroem2012}, i.e., when $\log (T_\text{eff}[\text{K}])>3.9$ and the surface hydrogen mass fraction is above $30\%$. For models where we apply the \Bes{} prescription, we use their recommended domain of $T_\text{eff}>30\,000\,\text{K}$ and the surface hydrogen mass fraction above $10^{-5}$, since this satisfies the assumptions of the Eddington stellar model and includes WNh stars, which are also adequately described by this prescription according to \citet{Bestenlehner2020}. Although the domains of these two mass-loss prescriptions are not equivalent, they both cover most of the MS, which is the longest phase and also the one during which we are interested in comparing different mass-loss rates.

\subsection{Post-main-sequence mass-loss rates}

\add{For parts of the evolution not covered by the aforementioned prescriptions, we use the same mass-loss scheme} as described in \add{the GENEC grids} by \citet{Ekstroem2012} \add{, which we briefly summarize here.} \add{RSG mass loss is calculated as follows, based on a fit of Fig.\ 3 in \citet{Crowther2001}:}
\begin{equation}
    \log\dot{M} = -13.83 + 1.7\log(L/L_\odot)\text{.}
\end{equation}

\add{The mass-loss rate of WR stars is computed with \citet{Nugis2000}, with an additional $Z$-scaling from \citet{Eldridge2006} based on the calculations of \citet{Vink2005}. For WN stars, the following equation is used:}
\begin{equation}
\begin{split}
    \log\dot{M} &= -13.6 + 1.63\log(L/L_\odot)+2.22\log (X_\mathrm{He}) \\
    &+ 0.85\log(Z_\mathrm{ini}/Z_\odot)\text{.}
\end{split}
\end{equation}
\add{For WC and WO stars, the mass-loss rate is as follows:}
\begin{equation}
\begin{split}
    \log\dot{M} &= -8.3 + 0.84\log(L/L_\odot) + 2.04\log(X_\mathrm{He}) + 1.04\log(Z) \\
    &+ 0.66\log(Z_\mathrm{ini}/Z_\odot)\text{.}
\end{split}
\end{equation}
\add{In the small domain covered by \citet{Graefener2008}, this prescription is used instead for WR stars:}
\begin{equation}
\begin{split}
    \log\dot{M} &= 10.046+\beta\log (\Gamma_\mathrm{e}-\Gamma_0) \\
    &- 3.5\log(T_\mathrm{eff}[\mathrm{K}]) + 0.42\log(L/L_\odot)-0.45 X_\mathrm{H}\text{,}
\end{split}
\end{equation}
with $\Gamma_0=0.326-0.301\log(Z/Z_\odot)-0.045\log(Z/Z_\odot)^2$, and\linebreak $\beta =1.727+0.25\log(Z/Z_\odot)$.

\add{BSG mass loss is computed in the same way as MS OB stars, since their stellar parameters fall into the accepted validity domain of those prescriptions.}

\add{Outside the validity domain of all specialized mass-loss prescriptions, we apply the prescription by \citet{Jager1988}:}
\begin{equation}
\begin{split}
    \log\dot{M} = \sum_{j=0}^4 \sum_{i=0}^{5-j} a_{ij} T_i&\left(\frac{\log (T_\mathrm{eff}[\mathrm{K}])-4.05}{0.75}\right) \\  
    &\times T_j\left(\frac{\log(L/L_\odot)-4.6}{2.1}\right)\text{.}
\end{split}
\end{equation}
\add{where $T_j(x)=\cos (j\arccos x)$ are Chebychev polynomials, and $a_{ij}$ are fitted coefficients found in Table VI of \citet{Jager1988}. The definition of each prescription's validity domain results in the above formula effectively only being used for YSGs.}


\section{Evolution modeling}
\label{sec:modeling}

In order to investigate the effect of mass loss on the evolution of massive stars, we computed grids of evolutionary models using the Geneva stellar evolution code \citep[GENEC,][]{Eggenberger2008}. Specifically, we computed models with the following parameter grid:
\begin{itemize}
    \item Initial mass: 20, 25, 30, 40, 50, 60, 66, 73, 80, 85, 95, 105, 120 $M_\odot$,
    \item Metallicity: 0.014 (Solar), 0.006 (LMC),
    \item OB mass-loss prescription: \Vin{}, \Bes{}.
\end{itemize}

The full grid contains one model for each possible combination of these three parameters, resulting in a total of 52 models computed for this work. The set of initial masses was chosen to be roughly equidistant in the HRD at the zero-age MS (ZAMS), but having a slightly finer resolution in the higher mass range. The exact initial elemental abundances for the models at solar and LMC metallicity are described in \citet{Ekstroem2012} and \citet{Eggenberger2021}, respectively. All of the models are non-rotating in order to avoid introducing an additional source of uncertainty as well as new effects that may blur any conclusions pertaining to mass loss only. \add{Convection was treated using the Schwarzschild criterion and a step-overshoot scheme with overshooting parameter $\alpha_\mathrm{ov}=0.1$.} For all other physical ingredients not explicitly mentioned here, we use the same setup as described in \citet{Ekstroem2012}. The models were computed until the core carbon mass fraction drops below $10^{-5}$ at the end of core carbon burning, except both solar $40\,M_\odot$ models, as well as the solar $25\,M_\odot$ model using \Vin{}, for which convergence issues prevented the computation from finishing. These models will be excluded from analysis at the relevant points. 


\section{Early evolution and yields}
\label{sec:early-evolution}

\subsection{Characterization of mass-loss prescriptions}

\add{From the evolution models we obtain that, at} solar metallicity, \Vin{} produces a mass-loss rate that is systematically around one order of magnitude above \Bes{} across the entire initial mass range. At LMC metallicity, the difference between the two mass-loss rates is smaller, as the \Vin{} recipe scales \add{down} for \add{lower} metallicity, whereas our implementation of \Bes{} does not. Independent of their accuracy, the large contrast in the resulting mass-loss rate is beneficial for the purpose of comparing the effect of \add{weaker versus stronger} MS mass loss on stellar evolution.

\begin{figure}
    \centering
    \includegraphics{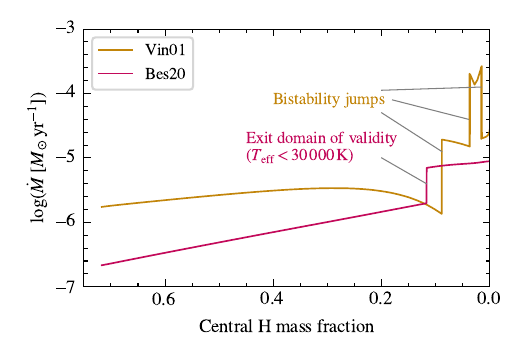}
    \caption{Evolution of the mass-loss rate during the MS of the $60\,M_\odot$ models at solar metallicity using the \Vin{} and the \Bes{} prescriptions.}
    \label{fig:mdot-t-MS}
\end{figure}

The \Vin{} mass-loss rate is much more variable during the MS for stars above 60 solar masses. This can be attributed to the fact that these more massive stars expand more during their MS, and thus the cooler atmospheres cause the mass loss to be boosted due to the bistability jumps. To illustrate the evolution of the mass-loss rate during the MS using the different prescriptions, we show an example in Fig.\,\ref{fig:mdot-t-MS}. As expected, \Bes{} produces a monotonic increase in the mass-loss rate of around one order of magnitude during the MS as the star steadily increases its luminosity and therefore its Eddington parameter. Close to the end of the MS, the star is too cool and thus we switch to the stronger mass-loss rate by \citet{Jager1988}.

\subsection{Main-sequence evolution}

\begin{figure*}
    \centering
        \includegraphics{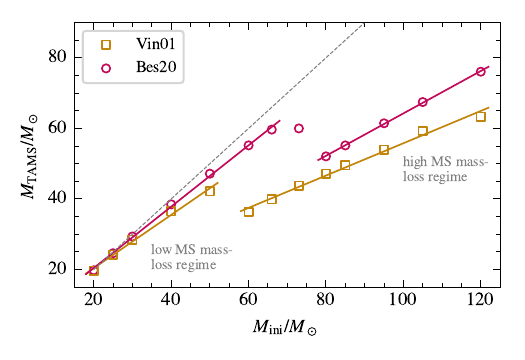}
        \includegraphics{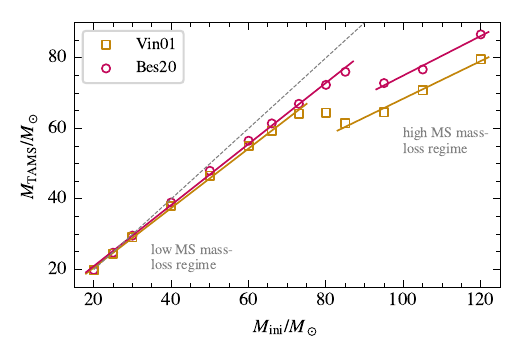}
    \caption{Total mass of the models remaining at the terminal-age main sequence (TAMS) as a function of initial mass at solar metallicity (\textit{left}) and LMC metallicity (\textit{right}). The data is fitted with a piece-wise linear function to indicate the two distinct mass-loss regimes. The gray dashed line indicates the 1:1 relation in both plots.}
    \label{fig:DeltaM_TAMS}
\end{figure*}

Fig.\,\ref{fig:DeltaM_TAMS} shows the mass remaining at the terminal-age main sequence (TAMS), defined here as the moment in time where the central hydrogen mass fraction of the star drops below $10^{-5}$. The results show a clear division into two distinct regimes of MS mass loss for both of the considered OB mass-loss prescriptions and at both metallicities. In each regime, the TAMS mass has a linear dependence on the initial mass, parametrized as $M_\mathrm{TAMS}=aM_\mathrm{ini}+b$, for which we determine the best fit parameters $a$ and $b$ using the Python routine \texttt{optimize.curvefit} provided by the \textit{scipy} package \citep{Virtanen2020}. The results are shown in Table \ref{tab:mtams-mini-fitting-params}, and drawn in Fig.\,\ref{fig:DeltaM_TAMS}. 

\begin{table}
    \centering
    \begin{tabular}{|c|cc|c|c|}
    \hline
    $Z$ & \multicolumn{2}{|c|}{OB $\dot{M}$ prescription} & $a$ & $b/M_\odot$ \\
    \hline
    $0.014$ & \Bes{} & ($M_\mathrm{ini}\leq 66 \,M_\odot$) & $0.869\pm 0.014$ & $3.1\pm 0.6$ \\
            &        & ($M_\mathrm{ini}\geq 80 \,M_\odot$) & $0.601\pm 0.007$ & $4.2\pm 0.7$ \\
            & \Vin{} & ($M_\mathrm{ini}\leq 50 \,M_\odot$) & $0.76\pm 0.04$ & $5.3\pm 1.3$ \\
            &        & ($M_\mathrm{ini}\geq 60 \,M_\odot$) & $0.458\pm 0.019$ & $10.0\pm 1.7$ \\
    \hline
    $0.006$ & \Bes{} & ($M_\mathrm{ini}\leq 85 \,M_\odot$) & $0.866\pm 0.012$ & $3.6\pm 0.7$ \\
            &        & ($M_\mathrm{ini}\geq 93 \,M_\odot$) & $0.56\pm 0.08$ & $19\pm 8$ \\
            & \Vin{} & ($M_\mathrm{ini}\leq 73 \,M_\odot$) & $0.844\pm 0.014$ & $3.6\pm 0.7$ \\
            &        & ($M_\mathrm{ini}\geq 85 \,M_\odot$) & $0.53\pm 0.05$ & $15\pm 5$ \\
    \hline
    \end{tabular}
    \caption{Best fit parameters for the piece-wise linear relation between TAMS mass and initial mass of the models at solar and LMC metallicity using the different OB mass-loss prescriptions. The parameters fulfill the relation $M_\mathrm{TAMS}=aM_\mathrm{ini} + b$.}
    \label{tab:mtams-mini-fitting-params}
\end{table}

\add{The two regimes apparent in Fig.\,\ref{fig:DeltaM_TAMS} arise from intrinsic differences in the mass-loss prescriptions and their validity limits. More massive stars have a higher mass loss and thus a larger core-to-envelope mass ratio. Thus further leads to a larger expansion during the MS and thus a cooler $T_\mathrm{eff}$.} Since there are multiple temperature thresholds which increase the applied mass-loss rate during cooling, such as the bistability jumps for \Vin{} or the limit of the validity domain for \Bes{}, MS expansion causes additional boosts in mass-loss rate for stars above \add{the initial mass corresponding to the regime break in Fig.\,\ref{fig:DeltaM_TAMS}. This means that the stars in the high MS mass-loss regime undergo most of their MS mass loss close to the end of the MS. For example, 64\% of the mass loss experienced during the MS by the solar $60\,M_\odot$ model using \Vin{} (plotted in Fig.\,\mbox{\ref{fig:mdot-t-MS}}) happens after the first bistability jump in the last 8\% of the MS evolution time.} \add{Stars at lower metallicity are generally hotter} and thus these stars must have a higher initial mass to reach the high mass-loss regime compared to solar metallicity. It is worth noting that for stars at the highest end of the mass range, the surface-temperature trend reverses because mass loss is so strong that it actually begins to expose hotter inner layers of the star. A star with an initial mass \add{of more than $100\,M_\odot$} already experiences intrinsically high mass loss for either of the OB mass-loss prescriptions without any boost from cooler effective temperatures.

Finally, it is worth drawing attention to another important effect of mass loss during the MS, namely the reduction of the central convective region. Stars with stronger mass loss have \add{less massive} helium cores by the end of their MS, implying that MS winds do not only remove hydrogen directly from the surface, but also prevent hydrogen within the star from being converted to helium. In other words, mass loss is effectively a damper of core nucleosynthesis despite being a surface process. 

\subsection{Stellar Yields}
\label{sec:yields}

There is much interest in exploring the chemical output of massive stars, given their unique ability to synthesize elements heavier than oxygen and expel them into the interstellar medium through their winds and their final SN explosion. While the aim of this paper is not to conduct a full study on the effect of mass loss on stellar yields, we present of some important implications of our results for yield calculations.

\begin{figure*}
    \centering
    \includegraphics{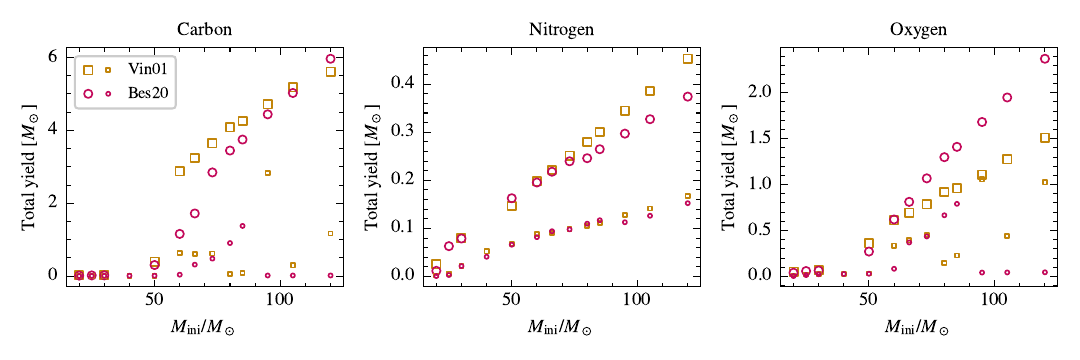}
    \caption{Total yields of carbon (\textit{left}), nitrogen (\textit{center}), and oxygen (\textit{right}) for models using different MS mass-loss prescriptions (\textit{colors} and \textit{shapes}) at solar metallicity (\textit{large}) and LMC metallicity (\textit{small}).}
    \label{fig:total-yields}
\end{figure*}

\begin{figure*}
    \centering
    \includegraphics{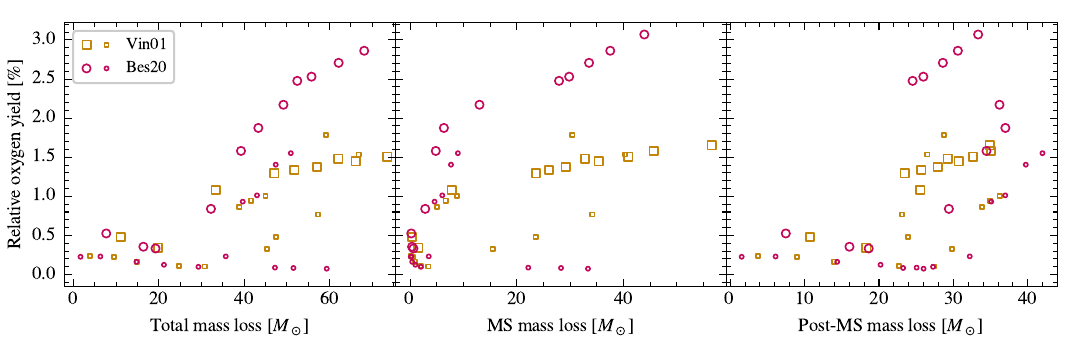}
    \caption{Relative yield (i.e.\ the ratio of total yield and total mass loss) of oxygen versus total mass loss (\textit{left}), MS mass loss (\textit{center}) and post-MS mass loss (\textit{right}) for models using different MS mass-loss prescriptions (\textit{colors} and \textit{shapes}) at solar metallicity (\textit{large}) and LMC metallicity (\textit{small}).}
    \label{fig:relyields}
\end{figure*}

Fig.\,\ref{fig:total-yields} shows the total yield of CNO elements for each of our models, calculated via

\begin{equation}
    \mathrm{Yield}_X = \sum X_\mathrm{surf, i}\, (M_\mathrm{i+1} - M_i)\text{,}
\end{equation}
where $M_i$ is the total mass of the star and $X_\mathrm{surf, i}$ is the surface mass fraction of element $X$ at evolution timestep $i$.

As expected, the results show a significant impact of metallicity on the yields of these elements. The separation by metallicity is especially apparent for the nitrogen yield, which is mainly dependent on the initial abundance of CNO elements and enhanced by the CNO cycle. For carbon and oxygen yields, we see a more prominent influence of the chosen MS mass-loss prescription than for the nitrogen yields. Since carbon and oxygen are expelled primarily during the late evolution of the star when the CO core becomes exposed, this illustrates how mass loss during the MS is able to impact the chemical structure of a star much later during the evolution. This chain of causality is indeed quite complex as there are multiple factors to consider. Firstly, the yields of carbon and oxygen are affected by the efficiency of the $^{12}\text{C}(\alpha,\gamma)\,^{16}\text{O}$ reaction during helium burning, which is set by the \add{core temperature during helium burning, which in turn is mainly a result of} the mass of the star at the end of the MS. Secondly, the transport of carbon and oxygen towards the surface depends on the timing of mass loss with respect to the interior evolution. If mass loss is \add{weaker}, core material has more time to be processed, but might not reach the surface before the end of the evolution and therefore not contribute to yields by winds. Conversely, \add{strong} mass loss might succeed in stripping the star down further, but the material will then not be enriched as much with carbon and/or oxygen. One may be tempted to try to relate the CO yields directly either to the strong post-MS mass loss (which strips the star), or to the weaker but longer MS mass loss (which sets the \add{TAMS} mass). However, as it turns out, \add{the total yield of carbon and oxygen correlates neither with the amount of mass lost in the MS, nor with the mass lost after the MS, nor with the total mass lost,} as shown in Fig.\,\ref{fig:relyields} for oxygen. Instead, the detailed mass loss history \add{throughout all phases of evolution must be considered in order to understand} pre-collapse yields of massive stars. Mass loss during the MS plays a pivotal role through its ability to determine the overarching mass loss history of a star by influencing the duration and sequence of its post-MS evolutionary phases. This will be explored further in \add{Sect.\,\ref{sec:populations}}.


\section{Implications for the evolution of Wolf-Rayet stars}
\label{sec:WR-evol}

\subsection{Spectral classification in evolution models}

\add{A WR star is defined as a spectral type characterized by strong and broad emission lines, which is an indicator of a strong, optically thick wind that obscures the hydrostatic layers of the star \citep{Abbott1987}.}
Stellar evolution models include only a simple approximated treatment of the stellar atmosphere, and thus offer no way of determining a spectral type directly. \add{Therefore}, stellar parameters such as surface abundances and temperature serve as proxies to spectroscopic definitions of the various classes of stars, although it is often unclear how closely this overlaps with observations.

In this work, we use these common criteria to study the consequences on stellar evolution and the perceived populations. \add{Our definitions of the various types of stars are given in Table \ref{tab:WR-subtypes}.} It should be noted that the WC/WO transition has been \add{redefined} since the analysis of WR stars by \citet{Georgy2012} from an abundance to a temperature criterion, due to new results from spectral modeling on evolutionary tracks by \citet{Groh2014a}. This has further been supported by recent analyses of WC and WO stars by \citet{Aadland2022}. Moreover, we have renamed the subtypes WNL and WNE in \citet{Georgy2012} to WNH and WN, respectively, in order to convey more directly that the differentiating parameter between these two subtypes is the surface hydrogen abundance, which does not necessarily correlate with the star's position in the HRD.

\begin{table}
    \centering
    \begin{tabular}{|c|c|c|c|c}
    \cline{2-4}
    \multicolumn{1}{c|}{} & Type & $\log(T_\mathrm{eff}\,[\mathrm{K}])$ & $X_\mathrm{H}$ & \multicolumn{1}{c}{}\\
    \cline{2-5}
    \multicolumn{1}{c|}{} & OB & \multirow{2}{*}{$\geq 4$} & \multirow{2}{*}{$> 0.3$} & \multicolumn{1}{c|}{MS} \\
    \cline{2-2}\cline{5-5}
    \multicolumn{1}{c|}{} & BSG & & & \multicolumn{1}{c|}{Post-MS} \\
    \cline{2-5}
    \multicolumn{1}{c|}{} & YSG & $3.7$ -- $4$ & \multicolumn{1}{c}{} & \\
    \cline{2-3}
    \multicolumn{1}{c|}{} & RSG & $\leq 3.7$ & \multicolumn{1}{c}{} & \\
    \cline{1-4}
    \multirow{4}{*}{WR}    & WNH & \multirow{2}{*}{$\geq 4$} & $10^{-5}$ -- $0.3$ & \multicolumn{1}{c}{}\\
    \cline{2-2}\cline{4-5}
        & WN & & \multirow{3}{*}{$\leq 10^{-5}$} & \multicolumn{1}{c|}{$\mathrm{C}\leq\mathrm{N}$}\\
    \cline{2-3}\cline{5-5}
        & WC & $4$ -- $5.25$ & & \multicolumn{1}{c|}{\multirow{2}{*}{$\mathrm{C}>\mathrm{N}$}}\\
    \cline{2-3}
        & WO & $\geq 5.25$ &  & \multicolumn{1}{c|}{}\\
    \hline
    \end{tabular}
    \caption{\add{Definitions of the types of stars used in this work.} }
    \label{tab:WR-subtypes}
\end{table}

\subsection{Surface hydrogen depletion}\label{sec:hdepletion}

\begin{figure}
    \centering
    \includegraphics{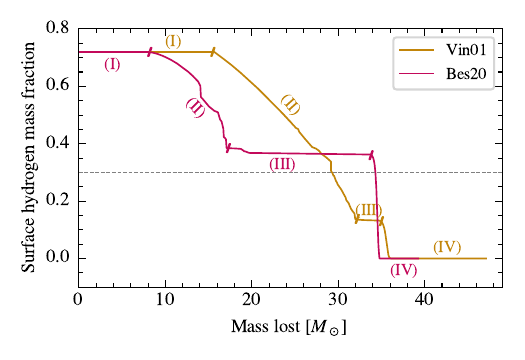}
    \caption{Evolutionary tracks of the $60\,M_\odot$ models at solar metallicity, showing the surface hydrogen mass fraction as a function of current total mass (hydrogen depletion curve). \add{The horizontal line indicates the threshold of surface hydrogen below which a hot star becomes classified as a WR in our models. The ``/'' markers on the tracks separate the curves into the four characteristic sections (I)--(IV), explained in the main text.}}
    \label{fig:hdepletion-m060z014}
\end{figure}

The WR definition based on low surface hydrogen abundance and high surface temperature intrinsically links them to a formation scenario involving significant mass loss, since hot and hydrogen-poor layers tend to be exposed by stripping. Let us therefore consider more closely the evolution of the surface hydrogen abundance $X_\mathrm{s}$ as a function of \add{total} mass loss $\Delta M$, and for simplicity refer to this relation as the \textit{hydrogen depletion curve} of a given stellar model. 

\add{In Fig.\,\ref{fig:hdepletion-m060z014}, we show this curve for our $60\,M_\odot$ solar metallicity models. As introduced in Sect.\,\ref{sec:early-evolution}, the \Bes{} model of this mass falls into the low MS mass-loss regime, whereas the \Vin{} model corresponds to high MS mass loss (See Fig.\,\ref{fig:DeltaM_TAMS}).} The curves can be divided into four phases, labeled as (I) to (IV). During phase (I), the star is stripping its chemically homogeneous envelope, and thus the surface hydrogen abundance remains constant. Phase (II) marks a gradual decline of surface hydrogen abundance, caused either by convective dredge-up after hydrogen-burning, or -- if the mass loss happens quickly enough -- by the exposure of the ZAMS convective core that has since shrunk but previously burned some of this region's hydrogen. After the end of the MS, when hydrogen shell-burning ignites around the core, a convective region is established above this shell which removes the chemical gradient in these layers. \add{This region is sometimes referred to as the ``intermediate convective zone'' (ICZ) in the literature, in reference to its position above the convective core but below the layers that eventually become the convective envelope in supergiants. When the ICZ is exposed,} the hydrogen depletion curve enters phase (III). Finally, when the hydrogen-burning shell is lost, the sharp chemical gradient below this shell causes the surface hydrogen abundance to drop to zero almost instantaneously and the star becomes a stripped helium star (phase IV).

The fact that \add{the two models in Fig.\,\ref{fig:hdepletion-m060z014}} yield different curves shows that there must be some feedback mechanism linking MS mass loss to the chemical structure of the interior layers before they are brought to the surface. \add{This affects in particular} the resulting length and hydrogen abundance of phase (III), i.e., the \add{ICZ}. If MS mass loss is high, the gas pressure and temperature in this region are low when hydrogen ignites in the shell since the envelope above it is smaller at that moment. Therefore, the \add{ICZ} will be smaller and more hydrogen-poor due to its proximity to the core \add{(cf.\ \Vin{} in Fig.\,\ref{fig:hdepletion-m060z014})}. Conversely, if the mass-loss rate during the MS is low, a large \add{ICZ} is established and is able to transfer hydrogen-rich material from higher layers into the interior of the star \add{(cf.\ \Bes{} in Fig.\,\ref{fig:hdepletion-m060z014})}. 

\subsection{Evolution of WR stars}
\label{sec:evol-wr}

\begin{figure*}
    \sidecaption
    \includegraphics{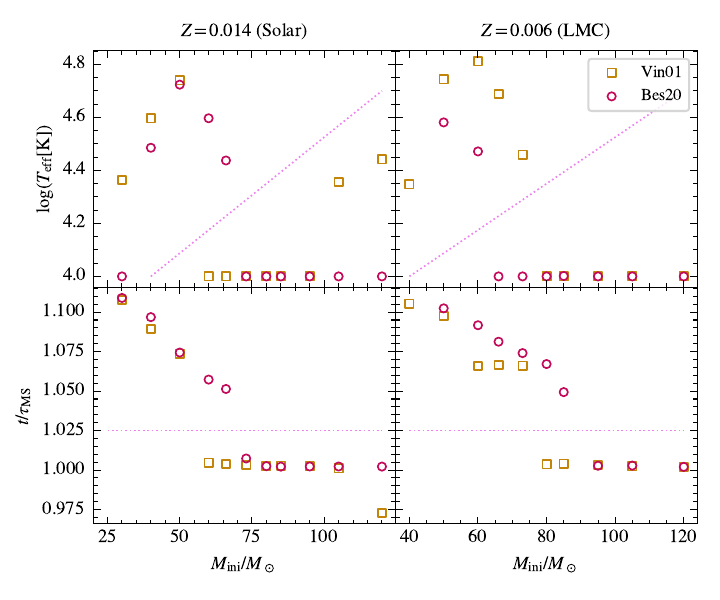}
    \caption{Effective temperature and evolution time as a function of initial mass at the point of WR formation, i.e.\ at the first computational time-step where each model fulfills our evolutionary criteria for WR stars. The colors and shapes indicate the chosen OB mass-loss prescription as either \Vin{} (\textit{yellow squares}) or \Bes{} (\textit{red circles}). The results are shown in separate panels depending on metallicity: solar (\textit{left}) and LMC (\textit{right}). The \textit{y}-axis of the bottom plots refers to the time-span elapsed since the ZAMS at the moment of WR formation, normalized by the time spent on the MS by each model. \add{The purple dotted line represents the separation of the models into two evolution scenarios, described in the text: the ``early-formed'' and ``late-formed'' WR stars are below and above this line, respectively.}}
    \label{fig:wr-formation-corner}
\end{figure*}

\begin{figure*}
    \sidecaption
    \includegraphics{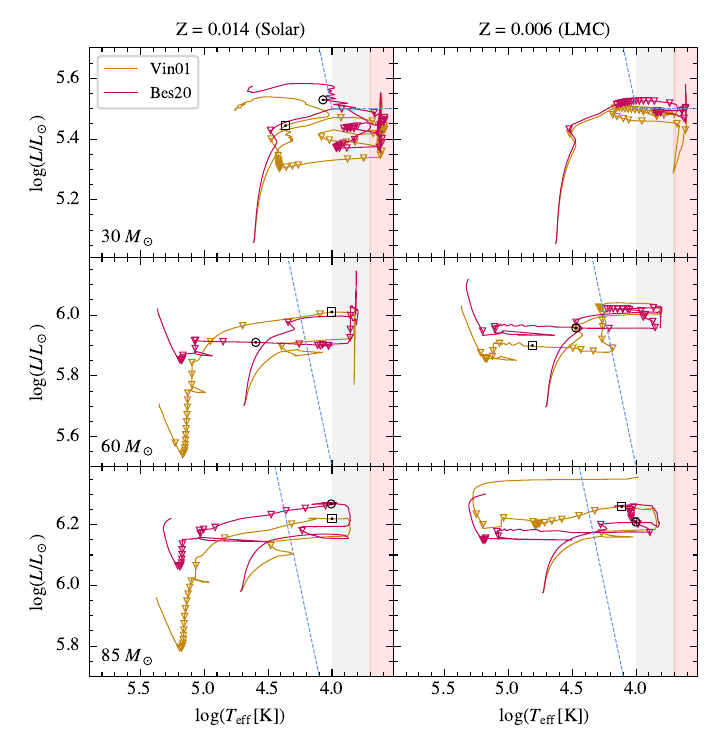}
    \caption{Evolutionary tracks in the Hertzsprung-Russell diagram (HRD) for the $30\,M_\odot$, $60\,M_\odot$, and $85\,M_\odot$ models (\textit{from top to bottom}) at solar metallicity (\textit{left}) and LMC metallicity (\textit{right}). The shaded red area indicates the domain of the red supergiant (RSG) phase ($\log (T_\mathrm{eff}\, [\mathrm{K}])< 3.7$), and the shaded gray area indicates the domain of the yellow supergiant (YSG) phase ($3.7\leq \log (T_\mathrm{eff}\, [\mathrm{K}])< 4.0$), as defined in GENEC for the application of specific mass-loss rates. The triangles mark time-steps of $20\,000\,\mathrm{yr}$ starting from core helium burning. The onset of the WR classification is marked by dotted squares and circles, for models using \Vin{} and \Bes{}, respectively. \add{The blue dashed line indicates the Humphreys-Davidson limit \citep{Davies2018b, Lamers1988}}}
    \label{fig:all-hrds}
\end{figure*}

We now characterize the properties of our models at the moment they enter the WR classification according to our defined criteria, to understand how stars in single-star evolution and population synthesis models transition to the WR phase. From the computations, all models at solar metallicity with an initial mass of at least $30\,M_\odot$ reach the WR classification, regardless of the applied MS mass-loss prescription. For LMC metallicity, the minimum initial mass for WR formation is $40\,M_\odot$ using the \Vin{} prescription, and $50\,M_\odot$ with \Bes{}. In Fig.\,\ref{fig:wr-formation-corner} an overview of the effective temperature and the evolution time of the models at the entry into WR classification is given as a function of initial mass, separated by metallicity. \add{The $y$-axis on the second row of this figure is the evolution time (since the ZAMS) normalized by each model's MS timescale. The MS therefore ends at $t/\tau_\mathrm{MS}=1.0$, and helium burning (lasting $\sim$$10\%$ of the MS timescale) ends at around $t/\tau_\mathrm{MS}=1.1$.}

\add{Fig.\,\ref{fig:wr-formation-corner} essentially reveals two types of evolution towards WR classification, which directly correspond to the two MS mass-loss regimes established in Sect.\,\ref{sec:early-evolution}. The complete evolution of these two different classes of stars in the HRD is shown for a few examples in Fig.\,\ref{fig:all-hrds}, which also shows the observational Humphreys-Davidson (HD) limit taken from \citet{Lamers1988}}\addr{, with the cool LMC component taken as $\log L = 5.5$ from \citet{Davies2018b}.}

\subsubsection{High MS mass-loss regime: ``early-formed'' WR stars}

\add{Stars in the high MS mass-loss regime are stripped very rapidly during the cooler YSG phase, having already lost most of their envelope during the MS. Since these stars have no hydrogen-rich convective zone, complete envelope stripping is not required in order for the stellar surface to become hydrogen-poor, and thus become classified as WR as soon as they cross the $\log (T_\mathrm{eff})$ threshold of $4.0$. Since this happens very shortly after the end of the MS (cf.\ Fig.\,\ref{fig:wr-formation-corner}), we refer to these WR stars as ``early-formed''. As they still retain a small amount of hydrogen on their surface, they are first categorized as subtype WNH.}

\add{Of the models depicted in the HRD (Fig.\,\ref{fig:all-hrds}), the ``early-formed'' (i.e., high MS mass-loss regime) WR stars are the $85\,M_\odot$ models at solar metallicity (both prescriptions) and LMC metallicity (\Vin{} only), as well as the $60\,M_\odot$ model at solar metallicity using \Vin{}.}

\add{In the HRD, such a star enters the WR phase while still having the radius and luminosity of a supergiant, \addr{since it still has an inflated envelope at this stage}. As it loses more mass, the star gradually transitions to hotter effective temperatures while reducing its luminosity. Once it has completely lost its hydrogen, it becomes a stripped WN star with an effective temperature hotter than $100\,000\,\mathrm{K}$. At this point, the star continues to strip down completely, losing the helium envelope above the convective helium burning core (which briefly expands just before being ejected causing a small redward spike in the HRD), and part of the carbon-enriched interior (at which point it would be classified as a WO/WC star). During this process, the star remains at a relatively constant high temperature, but significantly reduces its luminosity as it progressively reduces in size to less than one solar radius. The dimming is halted by the ignition of helium shell burning, which gives the star a burst of luminosity, further increasing its surface temperature. This creates the hook-like feature at the end of the HRD tracks which is characteristic of ``early-formed'' WR stars as well as some of the earliest (most massive) late-formed WR stars (e.g., $60\,M_\odot$ LMC models). The condition for this WR hook to appear is that the star needs to be stripped of hydrogen before helium shell burning begins, i.e.\ the mass loss needs to happen fast enough relative to the nuclear timescale. The more of the helium envelope is then stripped before helium shell burning begins, the lower the luminosity at the WR hook onset will be (e.g., compare both solar $60\,M_\odot$ models).}

\add{To summarize, stars in the high MS mass-loss regime reach the WR stage early and follow the evolutionary trajectory O--[YSG]--WNH--WN--WC/WO, where the YSG phase is a very short transition period, if it exists at all. Since for these stars, the onset of the WR stage (i.e., their hydrogen-depletion) coincides with the beginning of core helium burning, they are close to the ``Helium main sequence''. These ``Helium stars'' are used as a proxy to study WR evolution independent of the formation scenario \citep[e.g.,][]{McClelland2016, Yoon2017}. As we will discuss below, this proxy is only valid for the ``early-formed'' subclass of WR stars.}

\subsubsection{Low MS mass-loss regime: ``late-formed'' WR stars}

\add{The stars in the low MS mass-loss regime spend at least the first half of their helium burning time with a considerable envelope. This happens mostly in the cool region of the HRD, where \addr{strong RSG/YSG} mass loss gradually removes the envelope. Once their envelope is thin enough, these stars then transition to the blue region of the HRD. Since this envelope is hydrogen-rich due to the large \addr{ICZ that formed after the TAMS}, the stars continue their evolution being classified as BSGs until their envelope is completely stripped, the surface hydrogen abundance drops to zero, and the stars fulfill the criteria to be classified as WN.} 

\add{Of the depicted models in Fig.\,\ref{fig:all-hrds}, the ``late-formed'' (low MS mass-loss regime) WR stars are: the $85\,M_\odot$ LMC model using \Bes{}, both $60\,M_\odot$ LMC models, the $60\,M_\odot$ solar model using \Bes{}, and both $30\,M_\odot$ solar models. The $30\,M_\odot$ models at LMC metallicity as well as all $20\,M_\odot$ and $25\,M_\odot$ models lose too little mass to ever be considered WR and these finish their lives as RSGs.} 

\add{The ``late-formed'' WR stars follow the evolution path O--YSG/RSG--BSG--WN(--WC/WO), formally ``skipping'' the WNH stage and instead evolving directly as a hydrogen-free classical WR star with a more evolved core. Some of these stars can also have a pronounced BSG phase just after the MS before expanding to become YSGs/RSGs, which could be a result of our treatment of convection \citep[See also][]{Sibony2023}. Although these stars spend less time in the WR phase than the ``early-formed'' WR stars, they are more numerous due to their lower initial mass and therefore may also contribute significantly to observed and simulated populations of WR stars.}


\section{Core-collapse progenitors and stellar populations}
\label{sec:populations}

\subsection{Evolutionary endpoints}

\begin{figure}
    \centering
    \includegraphics{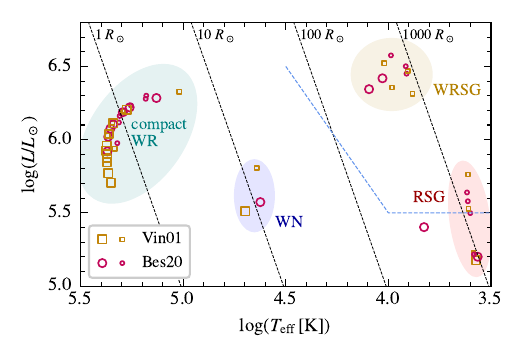}
    \caption{HRD showing the evolutionary endpoints of the models that were computed until the end of core carbon burning at solar metallicity (\textit{large markers}) and LMC metallicity (\textit{small markers}). The labeled ellipses show different populations of SN progenitors. Dashed black lines show equal radius of several different values. \addr{The blue line indicates the HD limit.}}
    \label{fig:end-hrd}
\end{figure}

\begin{figure}
    \centering
    \includegraphics{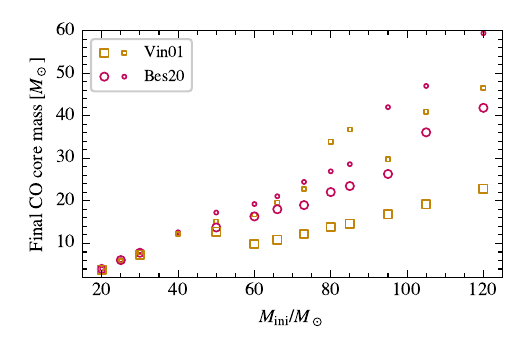}
    \caption{Core mass at the evolutionary endpoints of the models that were computed until end of core carbon burning at solar metallicity (\textit{large markers}) and LMC metallicity (\textit{small markers}). The core region is defined as the layers with a helium abundance below 1\%.}
    \label{fig:core-mass}
\end{figure}

We will now investigate the nature of the core-collapse progenitors by examining the physical characteristics of the models at their endpoint. Massive stars are commonly thought to end their lives as SNe, but recent studies now show that a considerable number of massive stars do not produce SNe but collapse silently or with a ``failed'' SN \citep{Smartt2009, Reynolds2015, Sukhbold2018, Beasor2023}. We computed (most of) the models until the end of core carbon burning, which we define as the point at which the central carbon mass fraction drops below $10^{-5}$. This happens at most a few years before core collapse. Considering the location of our models in the HRD, as shown in Fig.\,\ref{fig:end-hrd}, we see that our models split clearly into several populations of possible core-collapse progenitors:

\begin{table}
    \centering
    \begin{tabular}{|c|c|c|c|c|}
    \hline
         $M_\mathrm{ini}/M_\odot$ & \multicolumn{4}{c|}{Endpoint} \\
         \cline{2-5}
         & \multicolumn{2}{c|}{$Z=0.014$} & \multicolumn{2}{c|}{$Z=0.006$} \\
         & \Vin{} & \Bes{} & \Vin{} & \Bes{} \\
         \hline
         \hline
         20 & \multicolumn{4}{c|}{RSG} \\
         \hline
         25 & -- & YSG & \multicolumn{2}{c|}{RSG} \\
         \hline 
         30 & WN & WNH & \multicolumn{2}{c|}{RSG} \\
         \hline
         40 & \multicolumn{2}{c|}{--} & WNH & RSG \\
         \hline
         50 & \multicolumn{2}{c|}{WO} & \multicolumn{2}{c|}{WN} \\
         \hline
         60 & \multicolumn{4}{c|}{WO} \\
         \hline
         66 & \multicolumn{4}{c|}{WO} \\
         \hline
         73 & \multicolumn{4}{c|}{WO} \\
         \hline
         80 & \multicolumn{2}{c|}{WO} & WRSG & WC \\
         \hline
         85 & \multicolumn{2}{c|}{WO} & WRSG & WC \\
         \hline
         95 & WO & \multicolumn{2}{c|}{WC} & WRSG \\
         \hline
         105 & WO & \multicolumn{3}{c|}{WRSG} \\
         \hline
         120 & WO & \multicolumn{3}{c|}{WRSG} \\
         \hline
    \end{tabular}
    \caption{Estimated spectral type of the models at their evolutionary endpoint based on stellar parameters.}
    \label{tab:endpoints}
\end{table}

\begin{itemize}
    \item \textbf{Compact WR stars:} These are the endpoints of most of the models with strong enough mass loss. By the time of their death, most these stars have undergone so much mass loss that their surfaces are mainly composed of carbon and oxygen with only $\sim$20\% helium. These WC/WO stars \add{have removed almost all of the envelope above their helium-burning shell}, making them extremely hot and bright. The exception applies to the $50\,M_\odot$ LMC models, which are still formally classified as WN at the end of their evolution, but are also extremely hot ($\log (T_\mathrm{eff}\,[\mathrm{K}])> 5.25$).
    \item \textbf{``WR supergiants'' (WRSG):} This group contains the stars with the highest masses but comparatively weak mass loss, i.e.\ solar metallicity $105\,M_\odot$ and $120\,M_\odot$ using \Bes{}, as well as several high-mass LMC models. These stars have very \add{massive} cores which enable them to inflate their helium envelope after the onset of helium shell burning. They have the size and luminosity of supergiants while also having characteristics of WR stars such as the absence of hydrogen and enrichment of carbon and oxygen on the surface. They populate a \add{region considerably beyond the Humphreys-Davidson limit in} the HR diagram, and move there only within the last few years of their life, making their observation very unlikely if they do exist. However, one notable example of a star in this region of the HRD is $\eta$ Car \citep{Hillier2001}. It is presently unclear whether the derived HRD positions can be physically reached through the WR phase or whether this is just a consequence of the atmosphere treatment in GENEC.
    \item \textbf{WN stars:} These are the least massive WR stars in our sample (of those models that could be computed until the end), i.e.\ the $30\,M_\odot$ solar models as well as the $40\,M_\odot$ LMC model with \Vin{}. These stars have just lost their \addr{hydrogen envelope}, but still have an envelope with a surface made of the products of core hydrogen-burning, composed of mostly helium ($\sim 98$--$99$\%) and some nitrogen. Their helium-burning shell is still substantially covered, resulting in a cooler surface than the compact WR stars.
    \item \textbf{RSG stars:} These are the endpoints of the least massive stars, whose mass loss is too weak to significantly expose the hydrogen-burning shell and evolve back to the blue. They die as cool supergiants with only a slightly hydrogen-depleted surface which still consists of $50$--$65$\% hydrogen. The upper luminosity limit of RSGs in our models lies at around $\log (L/L_\odot)=5.8$, significantly higher than the upper limit observed among RSG progenitors of SNe, which lies closer to 5.2--5.3 \citep{Davies2020}.
\end{itemize}
The presumed spectral types for the endpoint of each model are summarized in Table \ref{tab:endpoints}.

It is also interesting to consider the mass of the core at the end of the computed evolution, since this quantity can be an important factor in driving the final SN explosion and determine the type and mass of the remnant left behind. The final CO core mass (defined as the mass of the region with a combined carbon and oxygen mass fraction above $95\%$) is shown in Fig.\,\ref{fig:core-mass} for the models of different metallicities and using different MS mass-loss prescriptions. It is evident that both metallicity and the MS mass-loss rate have a great influence on the final core mass of the star, especially for stars initially more massive than $50\,M_\odot$. The trend illustrates that stars at lower metallicity end with \add{more massive} cores (due to their higher compactness and lower mass loss). At any metallicity, stars with a \add{weaker} MS mass-loss rate also end with larger cores (with the exception of the $80\,M_\odot$ and $85\,M_\odot$ LMC models).

\subsection{Timescales and populations}
\label{sec:timescales}

\begin{figure}
    \centering
    \includegraphics{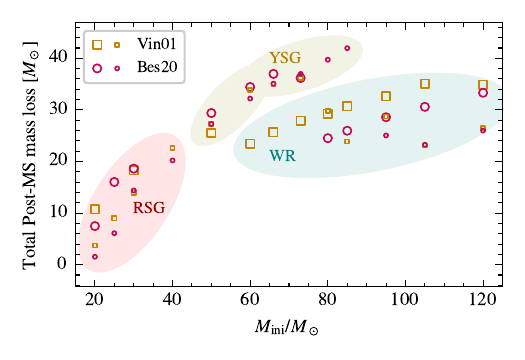}
    \caption{Post-MS mass loss by the end of core carbon burning of the models at solar metallicity (\textit{large markers}) and LMC metallicity (\textit{small markers}). The labeled regions denote which phase of mass loss is the dominant contributor to the total mass lost after the MS. The division of models into RSG/YSG and WR dominant post-MS mass loss corresponds exactly to the division into the low and high MS mass-loss regime, respectively.}
    \label{fig:endpoint-delM}
\end{figure}

\begin{figure}
    \centering
    \includegraphics{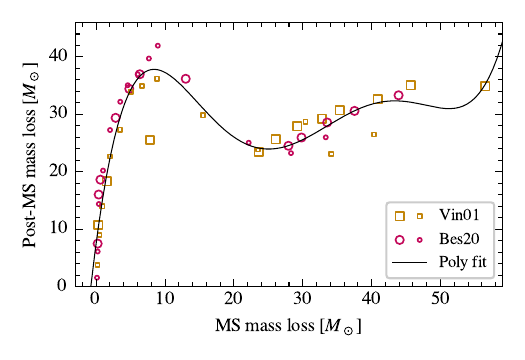}
    \caption{Mass lost after the MS versus during the MS for models at solar metallicity (\textit{large markers}) and LMC metallicity (\textit{small markers}). The data is fitted with a $n=5$ polynomial fit, which agrees with the data within 21.2\% (root mean square relative deviation). Visually, this illustrates that the mass loss experienced during the MS of massive star evolution is a reasonable predictor of the mass loss that will be experienced in the later stages of evolution, regardless of metallicity or MS mass-loss prescription. In particular, the shape of this curve illustrates the ``YSG bump'' for stars that lose around $10\,M_\odot$ during the MS, which spend the most time in the YSG phase out of any of the models.}
    \label{fig:msdelm-postdelm}
\end{figure}

\begin{figure*}
    \sidecaption
    \includegraphics{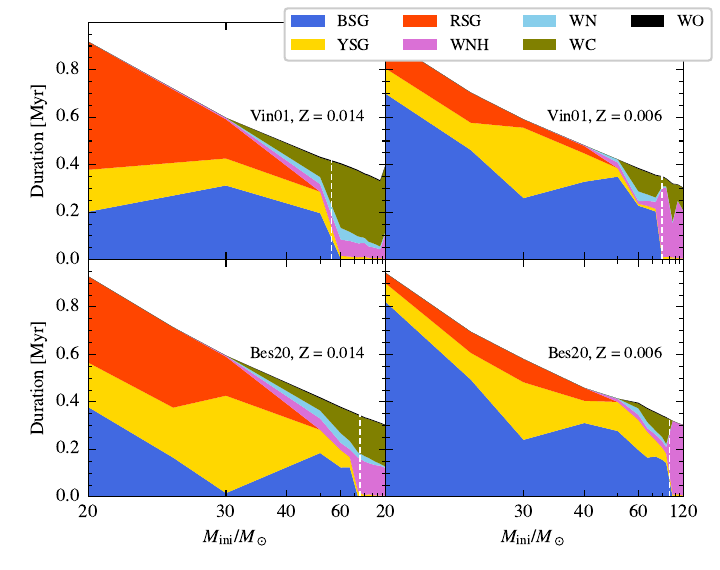}
    \caption{Duration of the various post-MS phases (colors), for each grid of models using the MS mass-loss prescriptions by \Vin{} (\textit{top row}) and \Bes{} (\textit{bottom row}), at solar metallicity (\textit{left column}) and LMC metallicity (\textit{right column}). The dashed white vertical lines represent the approximate location of the break between the low and high MS mass-loss regimes. The durations are shown cumulatively, so there is no information about the order in which the different phases are traversed. Furthermore, the $x$-axis is \add{a power scale (where $x\sim M_\mathrm{ini}^{-1.35}$), designed in} such a way that the area of a region corresponding to one phase is proportional to the number of stars currently in that phase, assuming a population of constant star formation rate using the standard initial mass function by \citet{Salpeter1955}.}
    \label{fig:timescales}
\end{figure*}

As we discussed in Sect.\,\ref{sec:WR-evol}, our evolution models follow a different trajectory based on which MS mass-loss regime they are in. The sequence and duration of the post-MS evolutionary phases is then a main determiner of the overall mass loss history of a star, since each evolutionary phase is characterized by a specific mass-loss rate. The results regarding the stellar yields (see Sect.\,\ref{sec:yields}) revealed the importance of the detailed mass loss history versus simple total mass loss estimates. In Fig.\,\ref{fig:endpoint-delM}, we consider the mass lost over all post-MS phases of the models. Our models can clearly be divided into groups depending on which post-MS phase is the most dominant contributor to the total mass loss of the star. As it turns out, this division into WR-dominated and RSG/YSG-dominated mass loss corresponds exactly to the division into high and low MS mass-loss regimes, respectively, which establishes an interesting causal link between MS mass loss and post-MS mass loss. Indeed, Fig.\,\ref{fig:msdelm-postdelm} shows that within our framework, the mass loss experienced during the MS is a good predictor of the mass loss that will be experienced after the MS, as underlined by the goodness of the $n=5$ polynomial fit to the data:

\begin{equation}
\begin{split}
        \Delta M_\mathrm{postMS}/10M_\odot &= 0.039x^5 - 0.62x^4 + 3.6x^3 \\
        &- 9.1x^2 + 9.1x + 0.76\text{,}
\end{split}
\end{equation}
where $x=\Delta M_\mathrm{MS}/10M_\odot$.

Expectedly, RSG mass loss is most significant for the stars of the lowest mass in our sample. Stars with masses between $50\,M_\odot$ and $85\,M_\odot$ are dominated by YSG mass loss, if they are in the low MS mass-loss regime. Otherwise, the YSG phase is too short, and the post-MS evolution is almost exclusively spent in the WR phase. Fig.\,\ref{fig:endpoint-delM} also shows that the stars with the largest post-MS mass loss are not the most massive stars, but the stars with the longest YSG phase, since in our models, this phase has the strongest mass loss \citep{Jager1988}. This effect can also be seen as a ``YSG bump'' in Fig.\,\ref{fig:msdelm-postdelm}. An interesting consequence of this is that at the same initial mass (e.g., $80\,M_\odot$), a star at lower metallicity may lose more mass in total due to it spending more time in the YSG phase than a star which has a higher metallicity and consequently loses more mass while still on the MS, but less mass afterwards. Unfortunately, mass loss in the YSG region of the HRD is still among the most poorly understood.

To understand the effect of MS mass loss on the timescales of various phases more closely, consider Fig.\,\ref{fig:timescales} which shows stacked line graphs of the duration of the post-MS phases obtained from the models. The results nicely summarize some conclusions of the previous analysis. Firstly, regardless of the MS mass-loss prescription or metallicity, stars in the low MS mass-loss regime spend more time as YSGs, during which time they are removing their large hydrogen-rich envelope. Secondly, these stars also spend less time as WNH, since they transition quickly from BSG to hydrogen-poor WN when they remove their \add{hydrogen-rich ICZ}. Comparing the different mass-loss prescriptions with each other, we notice that the main difference lies in the mass limit between the high- and the low MS mass-loss regime, while the overall trends within the regimes remain the same. Finally, the figure also shows that higher metallicity environments seem to favor RSGs, which is \add{partly} due to the fact that the higher core-to-envelope ratios at the end of the MS cause these stars to expand faster through the BSG and YSG stages and reach the RSG stage earlier in their evolution.

From the timescales of the various phases, one can further infer the distribution of massive stars in a population, assuming a constant star formation rate as well as some initial mass function. The number of stars $N_i$ in the population predicted to be currently in a certain evolutionary phase $i$ is then directly proportional to the integral of the product of the initial mass function $\mathrm{IMF}(M_\mathrm{ini})$ and the duration of that phase $\tau_i(M_\mathrm{ini})$ over the initial mass range of the population:
\begin{equation}
    N_i = K\int\tau_i (M_\mathrm{ini})\;\mathrm{IMF}(M_\mathrm{ini})\;\mathrm{d}M_\mathrm{ini}\text{,}
\end{equation}
where $K$ is some proportionality constant that normalizes the IMF to a given star formation rate. Using $\mathrm{IMF}(M_\mathrm{ini})=M_\mathrm{ini}^{-2.35}$ from \citet{Salpeter1955}, we can absorb this term into a new $x$ coordinate with $x=M_\mathrm{ini}^{-1.35}$, and the result is:
\begin{equation}
    N_i\propto \int\tau(M_\mathrm{ini})\;\mathrm{d}x\text{,}
\end{equation}
i.e., the number of stars $N_i$ is proportional to the area of the region associated with phase $i$ in Fig.\,\ref{fig:timescales}, creating a nice visual representation of populations of massive stars with different MS mass-loss rates and different metallicities.

\begin{table}
    \centering
    \begin{tabular}{|c|c|c|c|c|}
    \hline
        &   \multicolumn{4}{|c|}{Relative abundance [\%]} \\
    \cline{2-5}
    Phase & \multicolumn{2}{|c|}{$Z=0.014$} & \multicolumn{2}{|c|}{$Z=0.006$} \\
    \cline{2-5}
        & \Vin{} & \Bes{} & \Vin{} & \Bes{} \\
    \hline
    \hline
    MS   &  89.9  & 89.8 & 90.1 & 90.2 \\
    RSG  &  3.2   & 3.0  & 1.0  & 0.9  \\
    YSG  &  1.8   & 3.5  & 2.0  & 2.1  \\
    BSG  &  3.5   & 2.4  & 6.1  & 6.4  \\
    WR   &  1.6   & 1.3  & 0.8  & 0.4  \\
    \hline
    WNH  &  0.3   & 0.5  & 0.4  & 0.3  \\
    WN  &  0.2   & 0.2  & 0.1  & 0.1  \\
    WC   &  1.1   & 0.6  & 0.3  & 0.1  \\
    WO   &  \multicolumn{4}{|c|}{$<0.05$} \\
    \hline
    \end{tabular}
    \caption{Relative abundances of the different phases in populations of massive stars between $20\,M_\odot$ and $120\,M_\odot$, assuming constant star formation rate, a power-law IMF from \citet{Salpeter1955}, and different MS mass-loss rates and different metallicities. The values are numerical results rounded to one decimal place.}
    \label{tab:pop-dist}
\end{table}

Computing this integral numerically from the models (using trapezoid integration), and normalizing to $100\%$, we obtain the distributions given in Table \ref{tab:pop-dist}. The results show a few interesting properties, which we list here in no particular order:
\begin{itemize}
    \item There are fewer BSGs produced at high metallicities, due to the tendency of hot evolved stars to be classified as WR instead. In the lower mass range, higher metallicity stars expand faster than at lower metallicity, transitioning through more quickly to the RSG phase after the MS.
    \item There are fewer RSGs in lower metallicity environments, because \add{these stars expand more slowly off the MS}. Additionally, RSGs at lower metallicity cannot be progenitors of WR stars.
    \item There are fewer WR stars in lower metallicity environments, with a higher proportion of them being WNH.
\end{itemize}


\section{Discussion}
\label{sec:discussion}

The results obtained in this study are based on evolutionary models of single massive stars at solar and LMC metallicity and without rotation. \add{We now discuss} the applicability of the results on domains that were not directly explored in the simulations\add{, and evaluate the limitations and the impact of this study and its conclusions in the context of stellar evolution.}

\subsection{Metallicity}

The presence of metals in the atmospheres of stars has the general effect of increasing the opacity. This induces larger radiation pressure and therefore larger mass loss. Often, a metallicity dependence is included in mass-loss prescriptions in the form of a power law $\dot{M}\sim Z^a$, as is the case for \Vin{} in our models, but also others such as \citet{Bjoerklund2023}. The metallicity dependence of \Bes{} is less direct. In this prescription, the mass loss depends on the electron scattering Eddington parameter, with a force-multiplier $\alpha$ that supposedly corrects for the contribution of metallic lines to the total opacity. For implementation in GENEC, we have used this prescription with parameters calibrated by \citet{Brands2022} on LMC stars. These stars have a metallicity of around half of the Sun's and therefore exhibit a mass-loss rate that is lower than galactic OB stars by a factor of around 1.4--1.8, assuming a metallicity scaling with an exponent of around 0.5--0.85 \citep{Kudritzki2000, Vink2001}. Since we have directly applied this prescription at solar metallicity without adjustment, we expect the resulting mass-loss rate to be underestimated by this factor. However, we do not expect this to impact the results significantly, since the transition between the low and high MS mass-loss regime does not depend on the absolute scaling of the mass-loss rate.

In general, we find the studied effects of mass loss to be qualitatively similar at LMC and solar metallicities, with the mass limits between the low and high MS mass-loss regimes simply being shifted at lower metallicity. It is reasonable to suppose that for a further metallicity reduction, single star evolution could no longer produce ``early-formed'' WR stars (due to the lack of a high MS mass-loss regime). If the metallicity is reduced even further, WR stars would no longer be a possible outcome of single star evolution at all. At this point, binary interactions are expected to be the only remaining path towards WR formation \citep[e.g.,][]{Shenar2020}.

\subsection{Binarity and rotation}

Binarity has long been studied as an important aspect of massive star evolution \citep{Vanbeveren1991, Podsiadlowski1992, Wellstein1999, Mason2009}. Most massive stars interact with a binary companion at some point during their lives \citep{Sana2012, Sana2013}, and as many as 30\% of massive MS stars are the product of binary mergers \citep[e.g.,][]{Mink2014}. The mass flow induced by binary interactions may dominate (by far) the effect of radiatively driven winds in several regimes, complicating the comparison of the results from the present study to real populations of massive stars. However, the effect of mass loss or accretion via binary interaction on the evolution of stars may be similar to those caused by stellar winds and could in theory be interpreted in the same way. For example, a massive star that experiences a binary stripping episode during the MS (case A mass transfer) may experience a similar subsequent evolution as a star that has lost the equivalent amount of mass through MS winds. Moreover, the mass removal in cool regime can be treated as a proxy for the primaries in systems that undergo case-B mass transfer, i.e., those transferring mass when the primary expands towards the RSG regime. 

In an effort to produce a synthetic WR population for the LMC, \citet{Pauli2022} calculated both single and binary evolution models. In their illustrated binary models with case-B mass transfer, the mass transfer essentially removes a large part of the hydrogen-rich envelope. After the significant mass removal, the models predict that the primary will spend most of its lifetime as a hot WR, hydrogen-poor star, very similar to our cases with high MS mass loss. \addr{Hence, the fate of such a primary marks a case of what we call an ``early formed'' WR in our single-star models. Yet, the time scales are not expected to be similar by default. Compared to the short time scale of the binary mass transfer, the inner structure of the single star models might be more evolved as \addr{even} our single star models \addr{in the high MS mass-loss regime} will usually spend some time in the RSG and YSG stage, albeit this still being very short compared to the total He-burning time (around $0.3$--$2.5$\% of core He-depletion). }
\add{One aspect not mapped in this analogy is the possibility to regain hydrogen, which exists in binary systems in the case of mass transfer back from the secondary onto the primary.} Hence, the level of the plateau in the hydrogen depletion curve can also be higher \add{for a post-interaction primary} than for single-star evolution. This \addr{would} also \addr{contribute to a} variety of hydrogen surface abundances for partially stripped stars, \addr{such as those observed by} \citet{Ramachandran+2023,Goetberg+2023}, where some seem to spend a considerable amount of their post-MS evolution at temperatures and luminosities similar to main-sequence OB stars, albeit with much lower current mass.

Binary interaction can also inﬂuence the rotational behavior of stars, especially in the event of a merger. The surface anisotropy of rotating stars introduces an entirely new layer of complexity into the treatment of mass loss, even for single stars \citep[e.g.,][]{Poe1987, Bjorkman1993, Mueller2014}, which ideally requires modeling in (at least) 2-dimensions. Rotation also aﬀects other aspects of stellar evolution such as convective mixing within the star \citep{Maeder2000}, which we identified to be affected by prior mass loss with crucial implications for the subsequent evolution. 

As covering the whole implications from binarity and rotation is far beyond the scope of the present study, the results from this work should therefore be regarded as an exploration of the physical mechanisms relating mass loss and stellar evolution.

\add{\subsection{Unexplored uncertainties in single star evolution models}}
\label{sec:uncertainties-discussion}

\add{With this study being focused on mass loss, various parameters describing convection in 1D stellar models were not explored in this study. Yet, many of our results follow from properties of convective zones in the star, in particular the ICZ. ICZs form upon ignition of the hydrogen shell after the MS in \textit{all} of our models, albeit they are more prominent in stars that have previously experienced lower mass loss (cf.\ Fig.\,\ref{fig:hdepletion-m060z014}). The appearance of ICZs is also a consequence of our choice of convection parameters. For example, if one varies the overshooting parameter or the semiconvective efficiency, one can reasonably obtain models that do not form ICZs \citep{Schootemeijer2019}. Additionally, our use of the Schwarzschild criterion for convection favors the formation of larger convective regions as opposed to models which implement the more restrictive Ledoux criterion \citep{Sibony2023}.} 

\add{If we chose \addr{convection parameters which inhibit the formation of large ICZs, more} models would instead follow the ``early-formed'' WR formation scenario, where WR classification would be reached before total hydrogen depletion. As a result, models would predict a greater number of ``cool'' WNH stars. Moreover, stars would spend more time in the RSG phase and perhaps even stars with initial masses above $40\,M_\odot$ could become RSGs, since less efficient convection favors redward evolution \citep{Schootemeijer2019, Sibony2023}.}

\add{Secondly, it is known that there are large uncertainties in the mass-loss rates of various phases of post-MS evolution. These types of mass loss and their impact on massive stars has been the subject of previous investigations, for example \citet{Yang2023, Beasor2021, Massey2023} for RSGs, \citet{Owocki2019, Smith2006} for eruptive LBV mass loss, and \citet{Higgins2021, Sander2023} for WRs.}

\add{The uncertainties in convective mixing and supergiant/WR mass loss most importantly affect the evolution \textit{after} the MS. Since the post-MS treatment in GENEC is fixed within the framework of our study, our results highlight the ways that varying the mass-loss rate only during the MS can affect stars in every stage of evolution and show that the uncertainties in MS mass loss are indeed not negligible. MS mass loss must therefore be viewed as an important initial condition in studies concentrating on post-MS massive stars.}\\

\subsection{\add{Impact of MS mass loss and} existence of the mass-loss regimes}

Our results show a clear separation of the models into two MS mass-loss regimes for both the \Vin{} and the \Bes{} mass-loss prescriptions, and many subsequent conclusions follow this divide. These results are expected to propagate into the predictions of stellar population synthesis codes, which widely rely on prototypical stellar evolution models such as these ones. \add{Based on the evolution models, the choice of MS prescription mainly has an impact on the initial mass at which the MS mass-loss regime break is located, and thus most importantly affects stars between $55$--$75$\,$M_\odot$ at solar metallicity, and between $75$--$95$\,$M_\odot$ at LMC metallicity. It is also clear that without any MS mass loss, all stars would behave as if they were in the low MS mass-loss regime (see 1:1 relation in Fig.\,\ref{fig:DeltaM_TAMS}) and thus would end in RSGs or late-formed WRs. MS mass loss is therefore crucial for the evolution of early-formed WR stars, and in particular can explain the occurrence of WR stars with hydrogen (WNH).}

\add{One} may question whether the clustering into the two \addr{MS mass-loss} regimes \addr{presented in Fig.\,\ref{fig:DeltaM_TAMS}} is real. \addr{These regimes are the result of boosts in the mass-loss rate as a star expands while on the MS, yet it debated whether MS stars truly behave this way.} \citet{Bjoerklund2023} argue with their hydrodynamically consistent wind models for the absence of the iron bistability jump found by \citet{Vink1999}. However, due to the constraints of their parameter space to stars smaller than $80\,M_\odot$, it is unclear whether their models would also show a mass-loss regime change at higher masses, in particular as the results are in conflict with theoretical predictions by \citet{Krticka2021} and \citet{Bernini2023}. The $\dot{M}(\Gamma_\mathrm{{e}})$ dependencies in the \Bes{} prescription are motivated by observational studies such as \citet{Bestenlehner2014} and \citet{Brands2022}. While this recipe does not include a discontinuity in itself, there is still a sharp regime change for the \Bes{} models caused by the star exiting the domain of validity with $T_\mathrm{eff}$ below 30\,000\,K and entering the region of the HRD where we apply stronger YSG winds (see Fig.\,\ref{fig:mdot-t-MS}). The high mass-loss rate of YSGs in our models is a consequence of applying the prescription by \citet{Jager1988}, which is a generic prescription not designed specifically for YSGs and merely used as a ``fall-back'' since no better prescription exists for this regime. However, YSGs are indeed associated with significant mass loss \citep{Koumpia2020,Bonanos2023,Kraus2023,Humphreys2023}, which motivates the use of this prescription.

\subsection{Observability}

In the study of massive stars, observational constraints of evolutionary models are often provided in the form of population surveys, with key indicators being the abundance ratios between different types of stars, e.g.\ WR/O, BSG/RSG, and so on \citep[see, e.g.,][]{Eggenberger2002, Neugent2019, Higgins2020, Wagle2020}. We present a visual representation of these abundance ratios in populations with constant star formation rate in Fig.\,\ref{fig:timescales}, \addr{with data in Table \ref{tab:pop-dist}.} \addr{Assuming no WR stars form below the minimum mass of our model grid ($20\,M_\odot$), our predictions can be compared fairly to properties of observed WR populations. For example, we expect the WC/WN ratio to be approximately $2.2$ (MW, \Vin{}), $0.85$ (MW, \Bes{}), $0.6$ (LMC, \Vin{}) and $0.25$ (LMC, \Bes{}) for our grids. The \Bes{} populations fit remarkably well with the observations of \citet{Neugent2019}, who find a WC/WN ratio of $0.83$ for the MW and $0.23$ for the LMC. This indicates that weaker MS mass loss, as tested by our implementation of \Bes{}, may be a viable approach in order to correctly reproduce the distribution of WR stars in stellar populations. Of course, the other uncertainties discussed in this paper may also have an impact on the WC/WN ratio.}

\add{Moreover, we note the discrepancy between our modeled evolutionary tracks (see Fig.\,\ref{fig:all-hrds}) and the observational Humphreys-Davidson (HD) limit in the Galaxy and the LMC \citep{Lamers1988, Humphreys1984}. Especially the models using \Bes{} spend much more time in the ``forbidden'' cool area of the HRD, which suggests that this mass-loss rate is lower than reality, or that there are other mechanisms involved in avoiding too much inflation, in particular for the more massive models. \addr{To cite an extreme example, the $85\ M_\odot$ LMC model in Fig.\,\ref{fig:all-hrds} spends $\sim$\,$250\,000$ years beyond the HD limit, which represents most of its helium-burning lifetime}. \addr{However, models with initial masses as low as $40\,M_\odot$, including those from previous work \citep[e.g.,][]{Ekstroem2012}, do also cross the HD limit.} \citet{Gilkis2021} has suggested that an appropriate choice of rotational velocities and convective overshooting parameters can satisfy the HD limit, but not without in turn creating an excess of hot supergiants. In general, the tendency of evolutionary tracks to go beyond the HD limit is a common problem in evolution modeling that is still a subject of active investigation \citep[e.g.,][]{Higgins2020}.}

As mentioned previously, the conclusions of this work should be taken as an analysis of the physical link between mass loss and evolution. The effects explored here could potentially be used as indicators for the validity of certain mass-loss prescriptions, but this cannot be done without more complete model grids, nor without accounting for other uncertainties in stellar models. Observations could also be used to compare and calibrate the mass-loss prescriptions directly so that their implementation in stellar evolution codes is more robust.

\subsection{Theoretical domain limits}

For this work, we have defined different types of stars (e.g.\ RSG, YSG, WR, etc.) via strict criteria on stellar parameters that are easily provided by a stellar evolution code. Specifically, we have defined WR stars by a threshold of surface temperature and hydrogen mass fraction. Based on this definition, we have found that the classification of a WR star as ``early-formed'' or ``late-formed'' depends directly on the hydrogen content of its \add{ICZ} with respect to the threshold value of 0.3. If this shell is hydrogen-rich, the WR star forms ``late'', whereas if \add{it} is hydrogen-poor, the WR star forms ``early''. In general, we would reach similar conclusions regarding the formation and evolution of WR stars if the threshold were chosen differently. Shifting the WR surface hydrogen threshold to a higher value \citep[e.g., 0.4 as it is sometimes defined,][]{Schootemeijer2018} would simply lower the mass limit between early-formed and late-formed WR stars. Early-formed WR stars would form earlier, while the remaining late-formed WR stars would form at about the same moment, since the drop from hydrogen-rich to hydrogen-depleted happens almost instantaneously when the hydrogen-burning shell is removed and thus the exact hydrogen threshold value does not matter. 

Yet, it is questionable whether a strict hydrogen and temperature threshold is a valid assumption at all. In reality, WR stars are a spectroscopic class, independent of surface hydrogen abundance, and evolutionary criteria are a simply proxy defined to match the spectroscopy as closely as possible. Problematic aspects of this classification have already been identified, for example the existence of hot hydrogen-poor stars with optically thin winds whose mass loss is significantly lower than classical WN stars \citep{Vink2017, Shenar2020a}. These kinds of stars would be classified as WR by our evolutionary criteria, although they show spectra that are distinct from WR stars. It is part of future work to investigate the connections between spectroscopic and evolutionary classification and establish a more consistent picture of massive star evolution from both perspectives.


\section{Summary and conclusions}
\label{sec:conclusion}

In this paper we performed a systematic study of the effects of MS mass loss on the evolution of massive stars in the range of $20\,M_\odot$ to $120\,M_\odot$ at two different metallicities. Using the stellar evolution code GENEC, we compared the effect of applying two different mass-loss prescriptions for OB-type stars. For the direct impact on the MS evolution itself, including the total MS mass loss, we obtain two distinct regimes of \textit{low} or \textit{high} MS mass loss.

Mass loss during the MS acts early and over long timescales. The results show that it affects the evolution in two ways: First, it leads directly to the removal of the outer layers of the star, beginning with the hydrogen envelope. This has the general consequence of exposing metals on the surface and increasing the effective temperature once enough material has been removed. In this regard, mass loss acts in the same way on the MS as during any other phase. However, MS mass loss has shown here to have a second effect which is unique to this phase and acts more indirectly, namely through its connection with the formation and evolution of interior convective zones. Our results show that mass loss not only affects the size of the convective core during the MS, but more importantly, MS mass loss is almost the sole determinant of the mass a star has when it inflates to the RSG stage, which arguably marks the most dramatic structure change in the life of a massive star. It is \addr{right before the inflation} that \addr{our models} form a hydrogen-burning shell with a convective zone above it \add{(known as the ICZ)}, and that the core contracts and begins to burn helium in the center. The size and composition of the \add{ICZ} is directly dependent on MS mass loss, and will go on to play an important role in determining the \addr{formation and appearance of WR stars. Their evolution across the HRD, which is mostly independent of our WR classification scheme, is also influenced by MS mass loss.} 

After the inflation, mass loss loses its connection to the evolution of the stellar interior and its effect becomes more straightforward, simply stripping the star progressively. A useful visual representation of how MS mass loss affects the interior structure of stars is the \textit{hydrogen depletion curve} (e.g., Fig.\,\ref{fig:hdepletion-m060z014}), which shows a ``fossilized'' record of the hydrogen abundance of interior layers as they were brought to the surface.

Our study shows that even during the MS, the effects of mass loss cannot be ignored and must be accounted for in the uncertainties of any stellar model.
One important conclusion of this paper is the importance of mass loss \textit{history} for the evolution of massive stars. This is shown in particular by the results regarding stellar yields, which do not seem to be correlated simply with either total mass loss or mass loss during just one phase of evolution. The MS plays a special role in determining the subsequent mass loss history of a star, since the mass loss during this phase determines the sequence and duration of the post-MS phases of evolution, each of which has its own characteristic mass-loss rate. The models considered here, which are prototypical of current population synthesis models, show the following possible evolutionary paths, which are \add{significantly influenced} by MS mass loss:
\begin{itemize}
    \item[(a)] OB $\longrightarrow$ RSG ($\longrightarrow$ blue loop)
    \item[(b)] OB $\longrightarrow$ RSG/YSG $\longrightarrow$ BSG $\longrightarrow$ WN ($\longrightarrow$ WC/WO)
    \item[(c)] OB $\longrightarrow$ WNH $\longrightarrow$ WN $\longrightarrow$ WC/WO
\end{itemize}
Stars with the lowest initial masses in our grid experience type (a) evolution due to their mass loss being too low to significantly strip the hydrogen envelope. Some of these stars may attempt to cross the HRD again, shortly becoming BSGs in a blue loop, but none will reach the WR phase. Type (b) evolution corresponds to the models that have strong enough mass loss to form WR stars, but are still located in the low MS mass-loss regime. As such, they spend a considerable amount of time as cool RSGs/YSGs and then evolve back to the blue and become hydrogen-depleted WN stars, and eventually nitrogen-depleted WC/WO stars if the mass loss is strong enough. These stars only enter the WR phase in the latter half of core helium burning, which we have referred to as ``late-formed'' WR stars. Finally, models with type (c) tracks evolve to the WR phase directly from the MS, only briefly passing through the YSG phase for a few thousand years. These models become WR stars close to the beginning of core helium-burning and we have therefore called them ``early-formed'' WR stars, representing WNH stars while hydrogen still remains on the surface. These evolutionary tracks correspond largely to the so-called ``Conti scenario'' of massive star evolution \citep{Conti1975, Maeder1996}, but the mass limits are very sensitive to the chosen \add{MS} mass-loss rate and metallicity. 

\addr{Next to our detailed study of MS mass loss, we have also addressed other uncertainties of the models in Sect.\,\ref{sec:uncertainties-discussion}, namely the parameters describing convection, which are critical for explaining the appearance of the ICZ.}

Although we only model the evolution of individual stars, \add{the obtained paths towards WR classification} can also be applied to binary systems with mass transfer. In systems where mass transfer occurs when stars enter the Hertzsprung gap, the mass loss onto the companion \add{fulfills a similar role as} the high assumed mass loss in the RSG and YSG stages, \add{albeit not necessarily in a similar amount of time}. Thus, the general distinction between stars entering the WR stage early, i.e., at the beginning of He burning or even sooner, or late remains valid even when accounting for binaries. To better predict the individual populations, more detailed studies for specific metallicities will be required.

 
\begin{acknowledgements}
  The authors thank the anonymous referee whose detailed and constructive comments helped improve the quality of this paper considerably. The authors acknowledge fruitful discussions made possible by the International Space Science Institute (ISSI) in Bern, through ISSI International Team project 512 (Multiwavelength View on Massive Stars in the Era of Multimessenger Astronomy, PI Oskinova). JJ acknowledges funding from the Deutsche Forschungsgemeinschaft (DFG, German Research Foundation) Project-ID 496854903 (SA4064/2-1, PI Sander) and is a member of the International Max Planck Research School for Astronomy and Cosmic Physics at the University of Heidelberg (IMPRS-HD). SE acknowledges the STAREX grant from the ERC Horizon 2020 research and innovation programme (grant agreement No. 833925). AACS is supported by the Deutsche Forschungsgemeinschaft (DFG - German Research Foundation) in the form of an Emmy Noether Research Group -- Project-ID 445674056 (SA4064/1-1, PI Sander) and acknowledges funding from the Federal Ministry of Education and Research (BMBF) and the Baden-Württemberg Ministry of Science as part of the Excellence Strategy of the German Federal and State Governments.
\end{acknowledgements}

\bibliographystyle{aa}
\bibliography{references}

\end{document}